\documentclass[
aps,%
12pt,%
final,%
notitlepage,%
oneside,%
onecolumn,%
nobibnotes,%
nofootinbib,
superscriptaddress,%
noshowpacs,%
centertags]%
{revtex4}
\RequirePackage{graphicx}

\def\refitem#1{\relax}
\def\lsim{\stackrel{\scriptstyle <}{\phantom{}_{\sim}}}

\begin{document}
\title{Viscosity and thermal conductivity effects at  first-order phase transitions in heavy-ion collisions}

\author{\firstname{D. N.} \surname{Voskresensky}}
\email{D.Voskresensky@gsi.de} \affiliation{ National Research
Nuclear University "MEPhI", Kashirskoe sh. 31, Moscow 115409,
Russia}
\affiliation{GSI Helmholtzzentrum f\"ur
Schwerionenforschung GmbH Planckstra$\beta$e 1, 64291 Darmstadt,
Germany}

\author{\firstname{V. V.} \surname{Skokov}}
\email{V.Skokov@gsi.de} \affiliation{GSI Helmholtzzentrum f\"ur
Schwerionenforschung GmbH Planckstra$\beta$e 1, 64291 Darmstadt,
Germany}

\begin{abstract}Effects of viscosity and thermal conductivity  on
the  dynamics of  first-order phase transitions are studied.
The nuclear gas-liquid and hadron-quark transitions in
 heavy-ion collisions are considered.
We demonstrate that  at non-zero thermal conductivity,
$\kappa \neq 0$, onset of spinodal instabilities occurs
on an isothermal spinodal line, whereas  for $\kappa =0$ istabilities
take place at lower temperatures, on an  adiabatic
spinodal.
\end{abstract}

\maketitle

\section{Introduction}
There are many phenomena, where first-order  transitions
occur between phases of different  densities. In nuclear physics
various first-order phase transitions   may take place
 in  the Early Universe,
 in heavy-ion collisions
 and in neutron stars (e.g., pion condensation,
kaon condensation,  deconfinement and chiral phase transitions),
see reviews ~\cite{Randrup,Shuryak:2008eq,Glendenning}.
At low collision energies the nuclear
gas-liquid (NGL)
first-order phase transition is
possible~\cite{Randrup,Ropke,SVB,D'Agostino:2005qj}.
At high collision energies  the hadron-quark gluon plasma
(HQGP)
first-order transition may occur (see e.g.
Ref.~\cite{Shuryak:2008eq}). Within a hydrodynamical approach
dynamical aspects of the
NGL and HQGP transitions were recently studied in
Refs.~\cite{SV1,SV2,RandrupGQP,SV3,R10}. An important role of
effects of non-ideal hydrodynamics  was  emphasized. In this talk, we review
some results obtained in Refs. \cite{SV1,SV2,SV3}.

\section{General setup}

We are interested in a description of long wavelength phenomena
at a first-order phase transition. Thereby  we accept an assumption that
the velocity of a fluctuation (seed) $\vec{u}$ is
much slower than the mean thermal velocity.
Then a description is possible in the framework of the standard system
of equations of non-relativistic non-ideal hydrodynamics:
the Navier-Stokes equation, the continuity equation,
and equation for the heat
transport:
\begin{eqnarray}
\label{Navier} mn\left[ \partial_{t} {u}_i + (\mathbf{u}\nabla)
{u}_i \right] &=& -\nabla_i P   + \nabla_k \left[ \eta \left(
\nabla_k u_i + \nabla_i u_k -\frac{2}{d} \delta_{ik} \mbox{div}
\mathbf{u}   \right)   + \zeta \delta_{ik} \mbox{div} \mathbf{u}
\right], \\
\label{contin}
\partial_{t}n +\mbox{div} (n \mathbf{u})&=&0, \\
 \label{therm}   T\left[\frac{\partial
s}{\partial t} +\mbox{div}(s\mathbf{u} )\right]&=&\mbox{div}(\kappa
\nabla T) +\eta \left(\nabla_k u_i + \nabla_i u_k -\frac{2}{d}
\delta_{ik} \mbox{div} \mathbf{u} \right)^2 +\zeta (\mbox{div}
\mathbf{u})^2\,,
\end{eqnarray}
$n$ is the density of the conserving charge (here the baryon
density), $m$ is the (baryon) mass, $P$ is the pressure, $\eta$
and $\zeta$ are the shear and bulk viscosities, $d$ is the
dimensionality of space, $T$ is the temperature, $s$ is the
entropy density, $\kappa$ is the thermal conductivity.

The simplest (yet non-trivial) example illustrating  principal features of a first-order phase transition in a mean-field
approximation is the Van
der Waals fluid.
The  adiabatic trajectories,  $\tilde{s}\equiv s/n \simeq$ const, for an expansion of a uniform fireball to vacuum,
are
shown in Fig. 1.  The
super-cooled vapor  (SV) and the overheated liquid (OL) regions
are between the Maxwell construction (MC)
and the isothermal spinodal (ITS)
curves, on the left and on the right respectively.
The adiabatic spinodal (AS) curve bounds the AS region from above.
For
$\tilde{s}_{cr}>\tilde{s}>\tilde{s}_{\rm MC2}$, where
$\tilde{s}_{cr}$ corresponds to the  value  of the specific entropy $\tilde{s}$
at the critical point and the line
with $\tilde{s}_{\rm MC2}$ passes through the point $n/n_{cr}=3$
at $T=0$, the system traverses the OL state (the region OL in Fig. 1), the ITS region (below the ITS line) and the AS region (below the
AS line). For $\tilde{s}>\tilde{s}_{cr}$ the system trajectory
passes through the SV state (the region SV in Fig. 1) and  the
ITS region.

All thermodynamic quantities can be expanded  near  an arbitrary reference point
$(n_{\rm r}, T_{\rm r})$, or  $(n_{\rm r}, \tilde{s}_{\rm r})$.
Considering the problem in $(n,T)$ variables it
is convenient to take the reference point in the vicinity of the critical point $(n_{cr} , T_{cr})$
but outside the fluctuation  (critical) region ($n_{cr}\mp \delta n^{\rm fl}$, $T_{cr}\mp
\delta T^{\rm fl}$)
assuming that the latter
 is very narrow.
Note that even in the fluctuation  region, the mean-field treatment can be used
provided one considers the system at  time scales, being shorter
than the scale responsible for a development of long-scale critical
fluctuations. Thus  we further put $(n_{\rm r}, T_{\rm r})=(n_{cr} , T_{cr})$.

Let us construct a generating functional in the variables $\delta n
=n-n_{cr}$, $\delta T=T-T_{cr}$, also called the Landau free energy, such that
$\delta (\delta F_L)/\delta(\delta n) = P - P_{f}+P_{\rm MC} $:
\begin{eqnarray}\label{fren}
&&\delta F_L = \int \frac{d^3 x}{n_{cr}}\left\{ \frac{cm[\nabla
(\delta n)] ^2}{2}+\frac{\lambda m^3 (\delta
n)^4}{4}-\frac{\lambda v^2 m(\delta n) ^2}{2}-\epsilon \delta n
\right\},
\end{eqnarray}
where $\epsilon = P_f-P_{\rm MC}$ is expressed through the
pressure at the MC.
The maximum of the quantity
$\epsilon$ is ${\epsilon}^{m}= 4\lambda v^3 /(3\sqrt{3})$.
For the sake of convenience, we will use $\epsilon$ normalized  to its maximal value,
$  \gamma_{\epsilon}  = | {\epsilon}/ {\epsilon}^{m}  |$ where
$0<\gamma_{\epsilon}<1$.
The first term in Eq. (\ref{fren}) is due
to the surface tension, $\delta F_{L,\rm surf}=\sigma S$, $S$ is
the surface of the seed.  For the Van der Waals equation of state:
\begin{eqnarray}\label{parame}
v^2 (T) =- 4 \frac{\delta{{T}} n_{cr}^2 m^2}{T_{cr}} ,
\quad\lambda_{cr} =\frac{9f_0 }{16}\frac{T_{cr}}{n_{cr}^2 m^3},
\quad \sigma =\sigma_0 \frac{|\delta
T|^{3/2}}{T_{cr}^{3/2}}\,,\quad  \sigma_0 =32 mn_{cr}^2T_{cr} c.
\end{eqnarray}

Using Eq.~(\ref{contin}), we rewrite Eq.~(\ref{Navier})
  in  the dimensionless
variables $\delta \rho =v \psi$, $\xi_i =x_i /l$, $i =1 ,\cdots ,
d$, ${\tau}=t/t_0$ as
 \begin{eqnarray}\label{dimens}
 &&- \beta \frac{\partial^2 \psi }{\partial
{\tau}^2} =\Delta_{\xi}\left(\Delta_{\xi}\psi +2\psi
 (1-\psi^2)+\widetilde{\epsilon}- \frac{\partial \psi}{\partial
 {\tau}}\right),\\
 &&l=\left(\frac{2c}{\lambda v^2}\right)^{1/2} ,\,\, t_0
 =\frac{2( \widetilde{d}\eta_{\rm r} +\zeta_{\rm r} )}{\lambda v^2
 \rho_{\rm r}},\,\,
 \widetilde{\epsilon}=\frac{2\epsilon}{\lambda v^3}, \,\, \beta
 =\frac{c\rho_{\rm r}^2 }{ ( \widetilde{d}\eta_{\rm r} +\zeta_{\rm
 r}
 )^2},\nonumber
  \end{eqnarray}
$\widetilde{d}={2(d-1)}/{d}$.
Only linear terms in the velocity $\mathbf{u}$ were kept in deriving this equation.
Since $v^2 \propto -\delta T$,
processes in the vicinity of the critical point are proven to be very
slow. This is known as the critical slowing down phenomenon.

Eq. (\ref{dimens}) should be supplemented by  Eq. (\ref{therm}) for the
heat transport, which owing to
Eq.~(\ref{contin})  after  linearization simplifies to
\begin{equation}\label{v-tS}
T_{\rm r}\left[
\partial_t \delta s
-s_{\rm r}(n_{\rm r})^{-1}
\partial_t \delta n
\right]=\kappa_{\rm r}\Delta \delta T.
\end{equation}
The variation of the temperature is related to the variation of
the entropy density $s[n,T]$ by 
$\delta T \simeq T_{\rm r} (c_{V ,\rm r})^{-1}\left(\delta s
-({\partial s}/{\partial n})_{T,\rm r}\delta n\right) ,$
where $c_{V ,\rm r}$ is the specific heat density.

Note that  Eq.~(\ref{dimens}) differs from the standard Ginzburg-Landau  equation
exploited   in  phenomenological
approaches.
The difference  disappears, if one sets   the  bracketed term in
the r.h.s. of Eq.~(\ref{dimens}) to zero.
Superficially this simplification   is legitimate,
if space-time gradients are small. However, for a seed
prepared in a fluctuation at $t=0$ with a distribution $\delta\rho
(t=0,\vec{r})$, the initial condition
$\frac{\partial \delta\rho (t, \vec{r}) }{\partial t}|_{t=0}\simeq
0$ should  be fulfilled. Otherwise, because of  a positive
kinetic energy contribution the probability of  fluctuations
is suppressed.
On the other hand, the above initial conditions
 cannot be simultaneously fulfilled, if a
differential equation is of the first-order with respect to time derivative.
Therefore,
there exists an initial stage of the dynamics of seeds ($t\lsim t_{\rm init}$), which  is not described by
the standard Ginzburg-Landau equation.

The time scale  for the relaxation of the density following Eq.
(\ref{dimens}) is $t_{\rho}\propto R$, where $R$ is the size of
a seed (as we will show below, at some stage  overcritical
seeds grow with constant velocity), and the time scale  for the
relaxation of the entropy/temperature, following (\ref{v-tS}), is
 \begin{eqnarray}\label{kT}
t_T = R^2 c_{V,\rm r} /\kappa_{\rm r} \propto R^2.
 \end{eqnarray}
The evolution of a seed  is governed by  the slowest mode. Thus,
for $t_T (R)<t_{\rho}(R)$, i.e for  $R<R_{\rm fog}$ ($R_{\rm fog}$
is  the typical seed size at which $t_{\rho}=t_T$), dynamics of
seeds  is controlled by Eq. (\ref{dimens}) for the density mode.
For seeds with sizes $R>R_{\rm fog}$, $t_T\propto R^2$ exceeds
$t_{\rho}\propto R$ and growth of seeds is slowed down. Thereby,
the number of seeds with the size $R\sim R_{\rm fog}$ may increase with
time. Estimates~\cite{SV1,SV2}
show that for the HQGP phase transition $R_{\rm
fog}\sim 0.1-1$ fm and for the NGL transition $R_{\rm fog}\sim
1-10$ fm $\lsim R_{fb}(t_{f.o.})$, where $R_{fb}(t_{f.o.})$ is the fireball size at the freeze out,
 $t_{f.o.}$ is the fireball evolution time till freeze out.
 Thus, thermal conductivity effects may manifest themselves  in heavy-ion
collision dynamics.

Note that seeds of a new phase are produced in an old phase
owing to   short-scale fluctuations. Fluctuations are
not incorporated in the above hydrodynamical equations defined in terms of
mean-field variables. Contributions of short-scale
fluctuations can be simulated  by a random force induced in Eqs.
(\ref{dimens}), (\ref{v-tS}), cf. \cite{PS}.

There are only two dimensionless parameters in Eq.
(\ref{dimens}),  $\widetilde{\epsilon}$ and $\beta$.
The parameter $\widetilde{\epsilon}$ is responsible for a difference
 between   the Landau free energies of the  metastable  and  stable
states. For $t_{\rho}\gg t_T$ (isothermal stage), $\widetilde{\epsilon}\simeq const$
and  dependence on the latter quantity disappears because of
$\Delta_{\xi}\widetilde{\epsilon}\simeq 0$.
 Therefore, dynamics is controlled only by the  parameter
$\beta$,
 which characterizes inertia (enters together with the
second derivative in time). This parameter is expressed in terms
of the surface tension and the viscosity as
 \begin{eqnarray}
\beta
 = (32T_{cr})^{-1}[\widetilde{d}\eta_{\rm r} +\zeta_{\rm r} ]^{-2}\sigma_0^2 m.
   \end{eqnarray}
{{The larger viscosity and the smaller surface tension, the
effectively more viscous (inertial)  is the fluidity of seeds.}}
For $\beta \ll 1$ one deals with the regime of effectively viscous
fluid and at $\beta \gg 1$, with the regime of perfect fluid.
For the NGL phase transition we estimate $\beta\sim 0.01$. For the
HQGP phase transition $\beta\sim 0.02-0.2$, even  for the very low
value of the ratio $\eta/s\simeq 1/(4\pi)$~\cite{SV1,SV2}.
Thus one deals with
effectively very viscous fluidity of density fluctuations in both
NGL and HQGP transitions.

\section{Dynamics of seeds in metastable area}
Let us  consider the stage $t_{\rho}\gg t_T$.  For an  expanding system
we have to  assume that typical time for the formation and evolution
of a fluctuation  of our interest $t_{form}+t_{\rho}$ is much
smaller than the typical fireball expansion time $t_{f.o.}$.
 Let us consider the situation, when   at very slow expansion with
$\tilde{s}(t)\simeq$ const  spatially quasi-uniform spherical
fireball of a large radius $R_{fb} (t)$ enters either the OL- state
or the SV- state (see  the corresponding curves in Fig. 1).  In case
$|n -n_{cr} |/n_{cr} \ll 1$, i.e. in the vicinity of the critical
point $(n_{cr},T_{cr})$,  solution (\ref{dimens}) describing
dynamics of the density in the fluctuation is presented  (in the
dimensional form) as \cite{SV1,SV2}:
 \begin{equation}\label{delr} \delta n (t,r)\simeq
\frac{v(T)}{m}\left[\pm\mbox{th} \frac{r-R_{n}
(t)}{l}+\frac{{\epsilon}}{2\lambda_{cr} v^3(T)}\right]+(\delta
n)_{cor},
\end{equation}
where the upper sign corresponds to the evolution  of  bubbles
and the lower one to the evolution of  droplets, $d=3$, and the
solution is valid for $|\epsilon/(\lambda_{cr} v^3(T))|\ll 1$.
 The  correction  $(\delta
n)_{cor}$ is responsible for the  exact baryon
number conservation. Considering $r$ in the vicinity of a
bubble/droplet boundary we get  equation describing evolution of the
seed size:
\begin{equation}\label{dim}
\frac{\beta t_0^2}{2}
\frac{d^2R_{n}}{dt^2}=\frac{3\epsilon}{2\lambda_{cr} v^3 (T)
}-\frac{2l}{R_{n}}-\frac{t_0}{l}\frac{d R_{n}}{dt}.
\end{equation}
Following this equation
 a bubble of an overcritical  size $R>R_{cr}=4l\lambda_{cr}v^3(T)/(3\epsilon)$
 of the stable gas
phase, or respectively a droplet of the liquid  phase, been initially
prepared in a fluctuation, will grow. On the early stage of the evolution  the size of
the bubble/droplet  $R_{n}(t)>R_{cr}$ grows with an acceleration.
Then it reaches a steady grow regime with a constant velocity
$u_{as}=\frac{3\epsilon l}{\lambda_{cr} v^3 (T) t_0}\propto
\gamma_{\epsilon}|T_{cr} -T|^{1/2}$.
 In the interior of the
seed $\delta n \simeq \mp v(T)/m$. The
correction $(\delta n)_{cor} \simeq v(T)R_{n}^3(t)/ (mR_{fb}^3)$
is very small for $R_{n}(t)\ll R_{fb}(t)$.

Substituting Eq.~(\ref{delr}) to Eq.~(\ref{v-tS}) for $T=$ const
we obtain
 \begin{eqnarray} \label{deltasinf}
&&\delta s = \left(\frac{\partial s}{\partial
n}\right)_{T}\left\{\frac{v(T)}{m}\left[\pm\mbox{th} \frac{r-R_{n}
(t)}{l}+\frac{{\epsilon}}{2\lambda_{cr} v^3 (T)}\right]+(\delta
n)_{cor}\right\}.
 \end{eqnarray}
While the temperature is constant in the interior
and exterior of the seed,  the entropy  and the density
are different  in the
interior and exterior regions. Nevertheless  the approximation of
a quasi-adiabatic expansion of the system  might be used
even, when the system reaches metastable region, provided the gas
of seeds is rare, or  $v(T)$ is small.

In Fig. \ref{EoS_dynamicsQ} we demonstrate numerical solutions of
the hydrodynamical equations in two dimensions ($d=2$) at the stage
$t_{\rho}\gg t_T$. We take $T/T_{cr}=0.85$ and compute the
configuration for $\eta \simeq 45$MeV$/\mbox{fm}^2$ and for $\beta
=0.2$ (effectively large viscosity). The choice $T_{cr}=162$~MeV,
$n/n_{sat}=1.3$ ($n_{sat}=0.16$ fm$^{-3}$) is relevant for the HQGP phase transition. We see
that undercritical seeds (disks) dissolve rather rapidly (typical time is $\sim$ several fm) but
overcritical seeds grow very slowly. Similar solutions exist
for bubbles. Therefore, one can hardly expect to observe
a manifestation of large size droplet/bubble remnants   in
 heavy-ion collisions.

The limit $\kappa =0$ is  specific. In case $|n -n_{P,max} |/n_{P,max}
\ll 1$, i.e. in the vicinity of the  point
$(n_{P,max},\tilde{s}_{P,max})$, corresponding solutions
of Eq.~(\ref{dimens}) describing dynamics of the density  can be presented in the form  (\ref{delr}) with the
only difference that $\delta T$ should be replaced by
$\delta\tilde{s}$, $\lambda_{cr}$ by $\lambda_{P,max}$ and $v(T)$
by $v(\tilde{s})$. Dynamics of $R_n (t)$ is determined by Eq.
(\ref{dim}), where one should replace values calculated at
$(n_{cr}, T_{cr})$ to the corresponding values at
$(n_{P,max},\tilde{s}_{P,max})$.
From   Eq. (\ref{dimens}) we obtain
$\delta\tilde{s}=0 =(\delta s n_{P,max}- s_{P,max}\delta
n)/n_{P,max}^2$
with $\delta n$ given by Eq. (\ref{delr}). From known values $\delta s$ and
$\delta n$ we can define  $\delta T$. Therefore, not only  the density, $n$, but
also the entropy density, $s$, and the temperature, $T$, vary  in
the surface layer and exhibit different values inside and outside a
seed. Contrary, the value $\tilde{s}$ remains constant.

\section{Instabilities in spinodal region}\label{Instabilities}

In this section the ``r''-reference point can be taken
arbitrary, therefore, we suppress the subscript   ``r''. To find solutions of
the hydrodynamical equations we put, cf. \cite{SV1,SV2},
\begin{eqnarray}\label{delns}
\delta n =\delta n_0 \mbox{exp}[\gamma t +i \mathbf{p}\mathbf{r}],
\quad \delta s= \delta s_0 \mbox{exp}[\gamma t +i
\mathbf{p}\mathbf{r}],
\quad T =T_{>}+\delta T_0
 \mbox{exp}[\gamma t +i \mathbf{p}\mathbf{r}],
\end{eqnarray}
where $T_{>}$ is the temperature of the uniform matter. From the
linearized equations of non-ideal hydrodynamics we find
 the increment,
$\gamma (p)$,
 \begin{eqnarray}\label{g2} &&\gamma^2 =
-p^2 \left[u_T^2 +\frac{(\tilde{d}\eta +\zeta)\gamma }{mn}+cp^2
+\frac{u_{\tilde{s}}^2 -u_T^2 }{1+\kappa p^2
/(c_{V}\gamma)}\right], \end{eqnarray}
 $u_T^2
=m^{-1}(\partial P/\partial n)_T$ and $u_{\tilde{s}}^2
=m^{-1}(\partial P/\partial n)_{\tilde{s}}$ are speeds of
sound at constant temperature and entropy, respectively.
Eq. (\ref{g2}) has three solutions. Expanding the solutions for
small momenta (long-wave limit) we find
\begin{eqnarray}
\gamma_{1,2} &=& \pm i u_{\tilde{s}} p + \left[ \frac{\kappa}{c_{V
}} \left( \frac{u_T^2}{u_{\tilde{s}}^2}-1\right)
 - \frac{\tilde{d}\eta +\zeta }{mn}
\right]\frac{p^2}{2},\label{solG1}   \\ \gamma_3 &=& -
\frac{\kappa u_T^2 p^2}{u_{\tilde{s}}^2c_{V }} \left[ 1 -
\frac{u_T^2-u_{\tilde{s}}^2}{u_{\tilde{s}}^2u_T^2}\left( c +
\frac{\kappa u_T^4 }{u_{\tilde{s}}^2 c_{V }^2}
   - \frac{(\tilde{d}\eta +\zeta) \kappa u_T^2}{mn
c_{V } u_{\tilde{s}}^2 } \right) p^2 \right]. \label{solG}
\end{eqnarray}
The solutions $\gamma_{1,2}$ correspond to the sound mode in the
long wavelength  limit, whereas $\gamma_3$ describes the thermal
transport mode. Below the ITS line (and above the AS line) $u_T^2
<0$, $u_{\tilde{s}}^2
>0$, solutions $\gamma_{1,2}$ correspond to an oscillation and damping, whereas  $\gamma_3$ describes an unstable
growing mode. Below the AS line, since there $u_{\tilde{s}}^2<0$
and $u_{T}^2<0$, the modes exchange their roles: the sound modes
become unstable, while the thermal mode is damped.

 For sufficiently high thermal conductivity and not as small $p$,
$\kappa p^2 /(c_{V }|\gamma|)\gg \nu$, $\nu =(u_{\tilde{s}}^2
-u_T^2 )/(-u_T^2),$  within the ITS region the most rapidly
growing mode
 ($\gamma \simeq
\gamma_{m}$, and $p\simeq p_{m}$) is the density mode:
\begin{eqnarray}\label{gmax}
\gamma_{m}=\gamma_{m}^{(1,2)} =\frac{(-u_T^2)mn_{cr}
 }{(2\sqrt{\beta}+1)(\tilde{d}\eta
+\zeta)},\quad p_{m}^2 =
\frac{(-u_T^2)\sqrt{\beta}}{(2\sqrt{\beta}+1)c}. \nonumber
\end{eqnarray}
  Using these
expressions we may rewrite the condition of a high thermal
conductivity  as $ \kappa\gg \nu c_{V } \sqrt{c}.$ Typical radius
of structures is $R_m\sim 1/p_m$, it decreases with increasing
$|\delta T|$.

The amplitudes of the temperature and density are related by
\begin{equation}\label{T0}
\delta T_0 =\delta n_0 \frac{T s [1-n(\partial s/\partial
n)_{T}/s]}{c_{V } n \left[1+{\kappa} /(\sqrt{c}c_{V } )\right]}.
\end{equation}
 Therefore,
the assumption of spatial homogeneity of  the system   fails right after  the ITS region is reached. An aerosol
(mist) of bubbles and droplets is formed for a typical time
$t_{aer}\sim 1/\gamma_{m}$.  For the Van der Waals equation of
state   we find that $\delta T_0 /\delta n_0
>0$, i.e. the temperature is larger in denser regions. However,
the amplitude of the temperature modulation is rather small
 for $
\kappa\gg \nu c_{V } \sqrt{c}.$

Let us consider the case of zero  shear and bulk viscosities and
non-zero, but  small thermal conductivity. For
$-u_T^2\ll 1$, i.e.  slightly below the ITS line,  we get
 \begin{eqnarray}\label{gam3m}
 \gamma_{m}= \gamma_{m}^{(3)} \simeq \frac{\kappa u_T^4}{4cc_V u_{\tilde{s}}^2}, \quad p_m^2 \simeq -u_T^2 /(2c)\,.
 \end{eqnarray}
 Therefore, in both considered  cases the instability occurs
at  the ITS line. However, for $\kappa \gg \nu c_V
\sqrt{c}$ the most rapidly growing mode corresponds to
$\gamma_m^{(1,2)}$ and in the opposite limit  $\kappa \ll \nu c_V
\sqrt{c}$, to $\gamma_{m}^{(3)}$.

 In Fig. \ref{waveamplQ}  we show
the time evolution of the density wave amplitudes given by the first
Eq. (\ref{delns}), for an undercritical value of $p$,
$p<p_{cr}=\sqrt{2}/l$ (left panel), and for an overcritical value (right panel) for
the same choice of the parameters $T/T_{cr}=0.85$, $T_{cr}=162$~MeV,
$n/n_{sat}=1.3$, as in Fig.  \ref{EoS_dynamicsQ}.  We see that the evolution is more
rapid compared to  the configuration presented in
Fig. \ref{EoS_dynamicsQ} and  the characteristic time scale $t_{ch}\sim
 10$~fm is comparable with the fireball expansion  time $t_{f.o.}$.
 Thus we may conclude that
 in heavy-ion collisions during expansion of the
fireball the system may linger in QGP phase
at $T< T_{cr}$. This means that the equilibrium value of the
critical temperature of the phase transition might be
significantly higher than the value which may be manifested  in
growth of fluctuations in experiments with heavy ions.
Fluctuations grow more
rapidly with decrease of $T$ below $T_{cr}$, $t_{ch}\sim 1.5 T_{cr}/|\delta T|$~fm.
Far from the critical point
 rapid grows of fluctuations  reminds effect of a warm
champagne.
There are prospects for observing specific
signatures of fluctuations  with a typical size, defined by $1/p_m$, in heavy-ion
collisions. Fluctuations of this kind might be  distinguishable from ordinary statistical fluctuations.

 The limit $\kappa =0$ is again specific.  Below the ITS line and above the AS line, the
thermal mode $\gamma_3$, which drives the system towards equilibrium for
a small  thermal conductivity, does not exist  for  $\kappa =0$.
Therefore, the evolution in the spinodal region is entirely governed by
 adiabatic sound excitations.
The increment $\gamma$ is given by Eq. (\ref{g2})  with $u_T$
 replaced by $u_{\tilde{s}}$:
 \begin{equation}\label{g1} \gamma^2 =
-p^2 \left[u_{\tilde{s}}^2 +\frac{(\tilde{d}\eta +\zeta)\gamma
}{mn}+cp^2 \right].
\end{equation}
Therefore, contrary to the case $\kappa \neq 0$, instability appears,
when the system trajectory crosses the AS line rather than the ITS
line. The value $u_{\tilde{s}}^2 =-\lambda_{P,max} v^2 (\tilde{s}
)=15\delta \tilde{s} T_{P,max} /(512 m)$ for the particular case
of the Van der Waals fluid in the vicinity of the point
$(n_{P,max}, T_{P,max})$ (taken as the reference point). This
result holds also in the case of   ideal hydrodynamics, where in
addition to  $\kappa =0$, the viscosity coefficients ($\eta$,
$\zeta$) are  zero.
From Eq.~(\ref{v-tS}), we find
\begin{equation}\label{s0}
 \delta s_0 =\delta n_0 s /n ,\quad \delta
T_0 =\delta n_0 \frac{T s [1-n(\partial s/\partial
n)_{T}/s]}{c_{V} n }.
 \end{equation}
 Thereby,  the temperature is modulated  similar to the
entropy density. For the Van der Waals equation of state we find
that $\delta T_0 /\delta n_0
>0$. The amplitude of the
 temperature modulation  is  larger than in case of  $
\kappa\ne0$, see Eq. (\ref{T0}).

Concluding,  for any  $\kappa\neq 0$ the solutions of Eq. (\ref{g2})
result in the onset of the instability already for  $u_T^2<0$ (i.e. below the ITS
line rather than below the AS line). Since in reality $\kappa$ is
indeed
nonzero, spinodal instabilities start to develop when the
trajectory crosses the ITS line rather than the AS,
i.e. at significantly higher temperatures. This
favors  an observation of signals of the spinodal decomposition in
the HQGP phase transition in heavy-ion collisions.

We expect that owing to a  manifestation of non-trivial fluctuation
effects (especially, of the spinodal decomposition at first-order
hadron-quark  transition)  a non-monotonous behavior of different
observables  as  function of collisional energy   may be observed.
This experimental analysis will be possible at RHIC, FAIR and  NICA.
Owing to properties of spinodal decomposition
a manifestation of specific structures with typical spatial size
may serve as a promising signal of the QCD  first-order phase transition.

{\bf{Acknowledgments.}} This work was supported in part by the
DFG grant WA 431/8-1.

\clearpage

\newpage

\begin{figure}[h]
\centering
\includegraphics[width=0.35\textwidth]{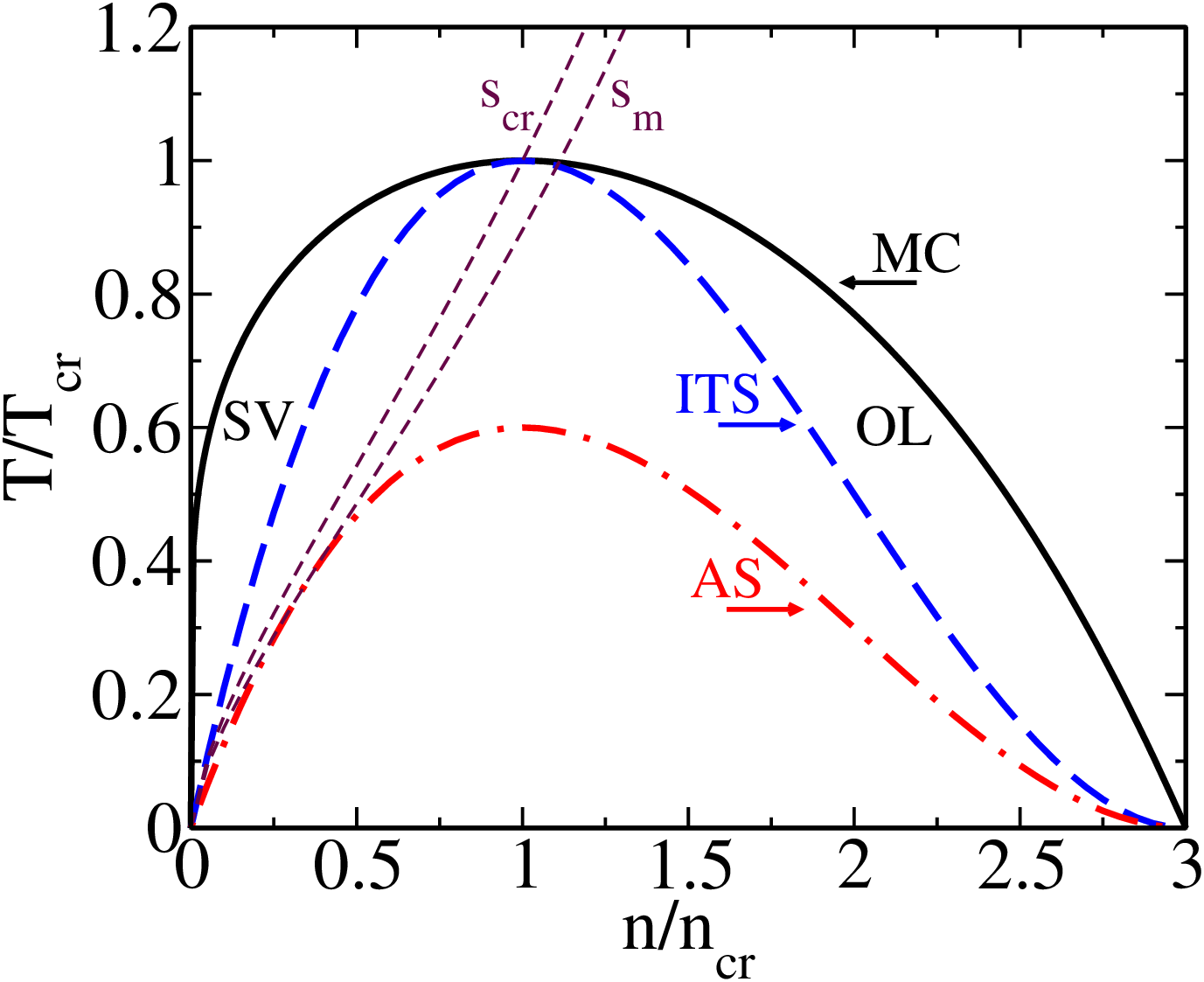}
\caption{The phase diagram of the Van der Waals equation of state,
$T(n)$-plane. The bold solid,  dashed and dash-dotted curves
demonstrate   the boundaries of the Maxwell construction, the spinodal
region at $T=$const and $\tilde{s}=$const, respectively. The short
dashed lines show adiabatic trajectories of the system evolution: the curve labeled
 $s_{cr}$ passes through the critical point; $s_m$, through the maximum pressure point $P(n_{P,max})$ on the
 $P(n)$ plane. }
\label{spinT}
\end{figure}
\clearpage
\begin{figure}
\centerline{%
{\includegraphics[height=5.0truecm] {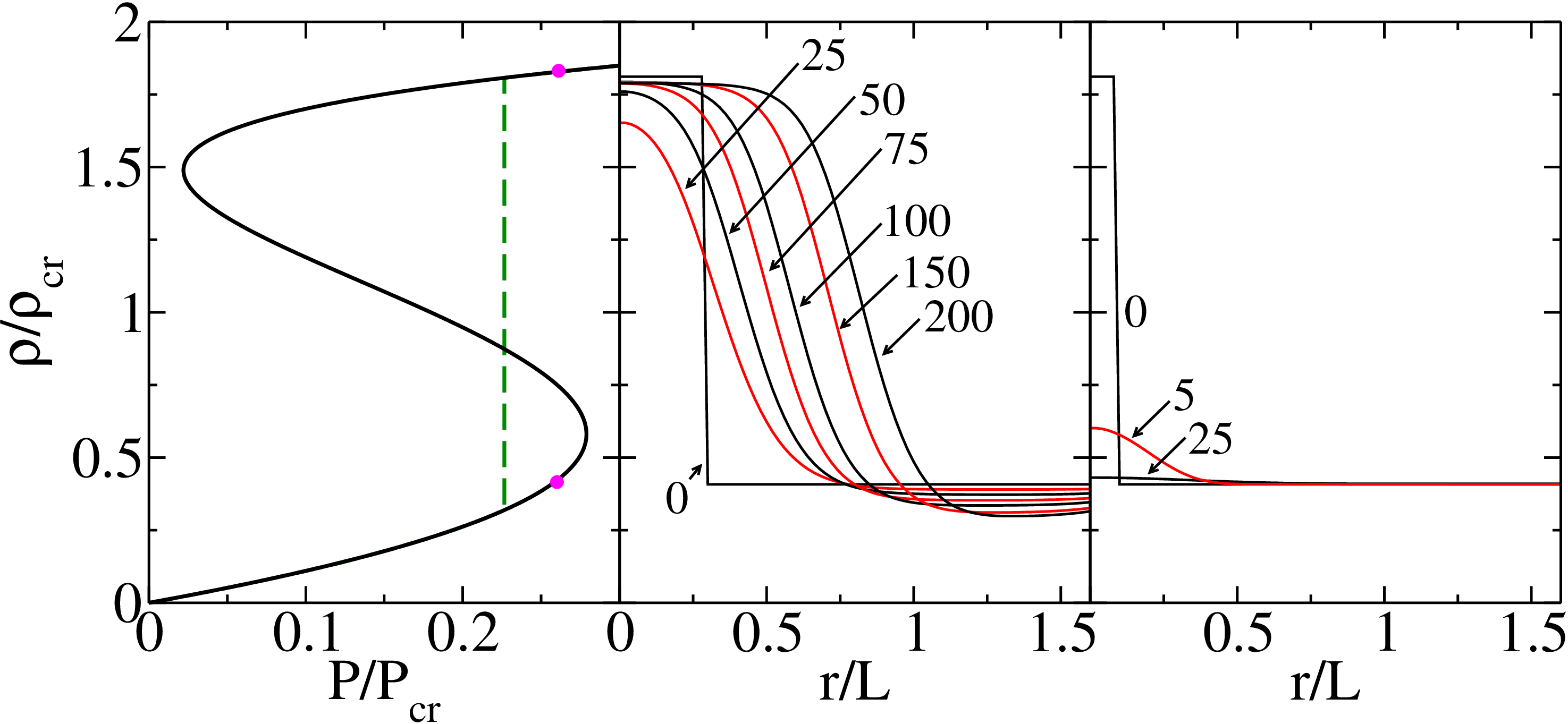}   } }
\caption{The isotherm for the pressure as
a function of the density, with initial and final configurations
shown by the dots  (left column).
 The dashed vertical  line shows  the
MC.
The initial state represents the stable liquid phase disk ($d=2$) in
metastable SV.
Middle column demonstrates the time evolution of the  density of
the overcritical liquid disk. The numbers near the curves (in $L$) denote time moments;
$r=\sqrt{x^2 +y^2}$, $|\delta {{T}}/T_{cr}|=0.15$, $L=
5$ fm. Right column, the same for the undercritical liquid
disk.
} 
\label{EoS_dynamicsQ}
\end{figure}
\clearpage
\begin{figure}\vspace{10mm}
\centerline{%
\includegraphics[height=4.0truecm]{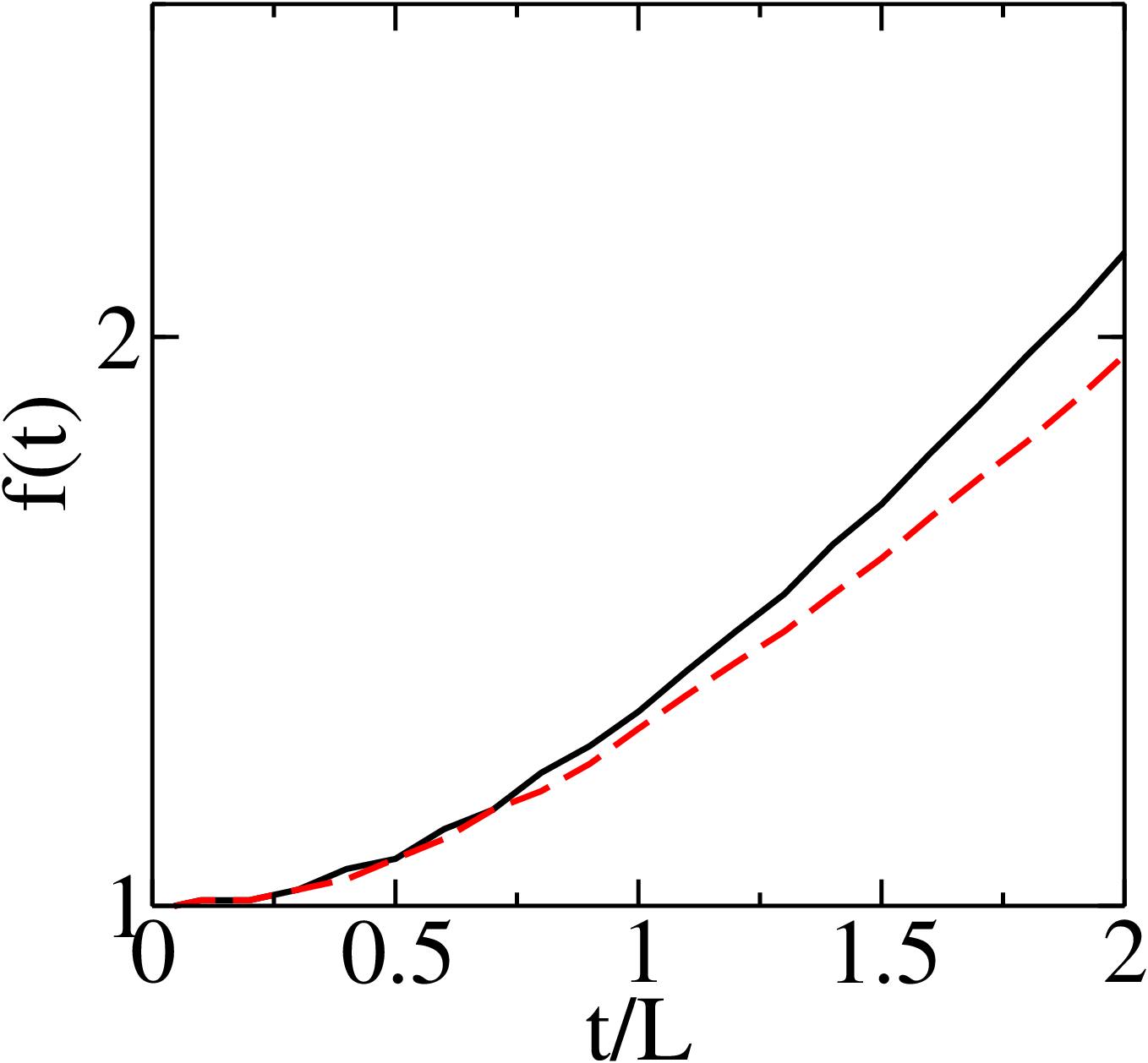}
\hspace{-1.0mm}
\includegraphics[height=4.0truecm] {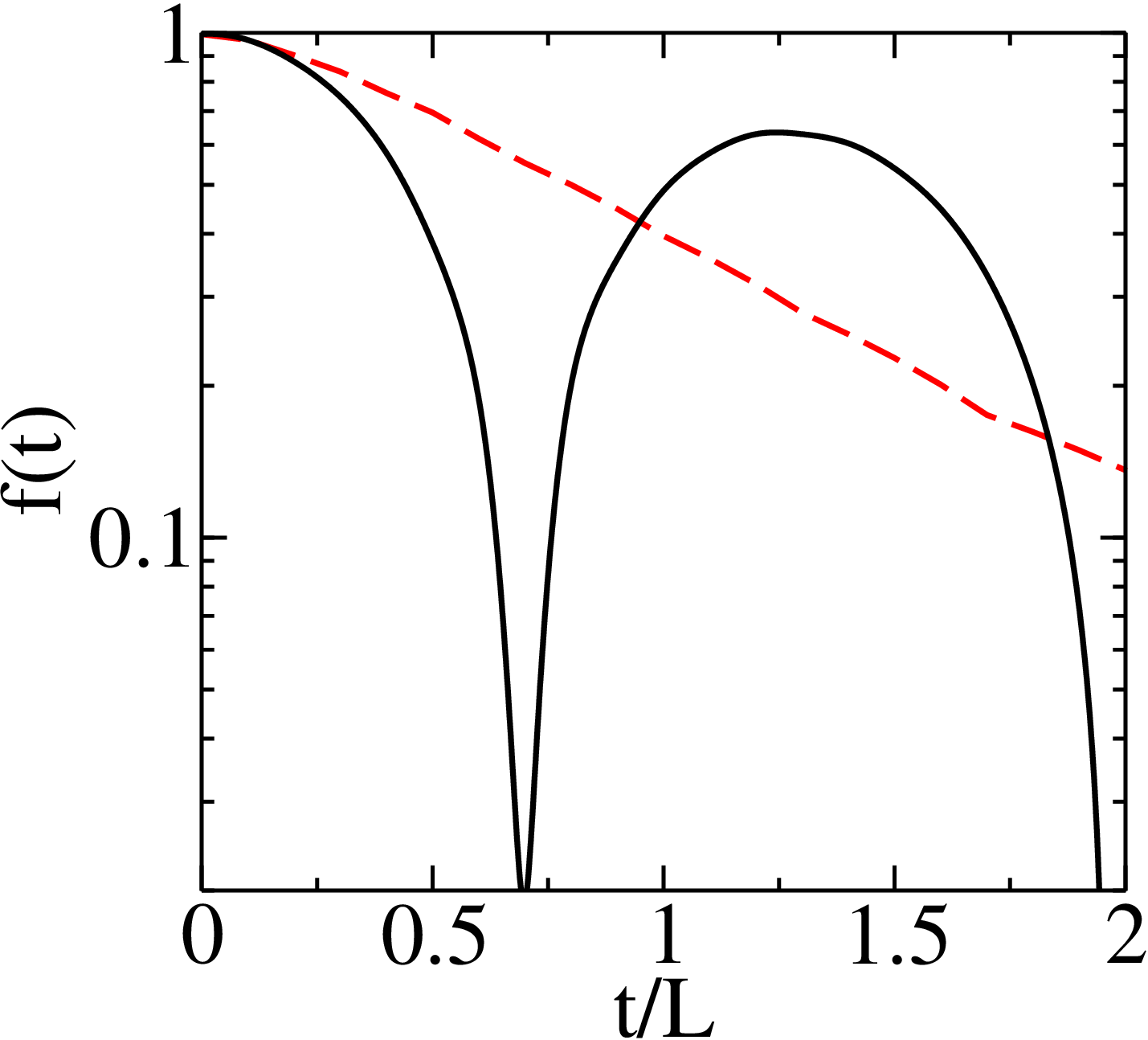} }
\caption{Time evolution of the wave amplitudes $f(t)$ defined as
$\delta n (t)$
normalized to the amplitude of the  initial disturbance. Solid
line  is for effectively small viscosity ($\beta =10$) and dash
line, for the large viscosity ($\beta =0.1$). Left panel: the
undercritical wave number $p=2/L$ (growing modes). Right panel:
the overcritical value $p=8/L$ (oscillation modes for large
$\beta$ and damped modes for small $\beta$).  Other parameters are
taken to be the same, as in Fig. \ref{EoS_dynamicsQ}.} \label{waveamplQ}
\end{figure}

\clearpage

\newpage

\begin{center}
FIGURE CAPTIONS
\end{center}
\begin{enumerate}

\item
{The phase diagram of the Van der Waals equation of state,
$T(n)$-plane. The bold solid,  dashed and dash-dotted curves
demonstrate   the boundaries of the Maxwell construction, the spinodal
region at $T=$const and $\tilde{s}=$const, respectively. The short
dashed lines show adiabatic trajectories of the system evolution: the curve labeled
 $s_{cr}$ passes through the critical point; $s_m$, through the maximum pressure point $P(n_{P,max})$ on the
 $P(n)$ plane.}

\item

{The isotherm for the pressure as
a function of the density, with initial and final configurations
shown by the dots  (left column).
 The dashed vertical  line shows  the
MC.
The initial state represents the stable liquid phase disk ($d=2$) in
metastable SV.
Middle column demonstrates the time evolution of the  density of
the overcritical liquid disk. The numbers near the curves (in $L$) denote time moments;
$r=\sqrt{x^2 +y^2}$, $|\delta {{T}}/T_{cr}|=0.15$, $L=
5$ fm. Right column, the same for the undercritical liquid
disk.
}

\item
{Time evolution of the wave amplitudes $f(t)$ defined as
$\delta n (t)$
normalized to the amplitude of the  initial disturbance. Solid
line  is for effectively small viscosity ($\beta =10$) and dash
line, for the large viscosity ($\beta =0.1$). Left panel: the
undercritical wave number $p=2/L$ (growing modes). Right panel:
the overcritical value $p=8/L$ (oscillation modes for large
$\beta$ and damped modes for small $\beta$).  Other parameters are
taken to be the same, as in Fig. \ref{EoS_dynamicsQ}.}

\end{enumerate}


\begin{thebibliography}{99}
\bibitem{Randrup}
P.~Chomaz, M.~Colonna, and J.~Randrup, Phys. Rep. {\bf 389}, 263
(2004).
\bibitem{Shuryak:2008eq}  E.~Shuryak,
arXiv:0807.3033 [hep-ph].

\bibitem{Glendenning}
N.~K.~Glendenning, Phys. Rep. {\bf 342}, 393 (2001).


\bibitem{Ropke} G.~R\"opke, L.~M\"unchow, and H.~Schulz, Phys. Lett. {\bf
B110}, 21 (1982).


\bibitem{SVB} H.~Schulz, D.~N.~Voskresensky, and
J.~Bondorf, Phys. Lett. {\bf B133}, 141 (1983).


\bibitem{D'Agostino:2005qj}
  M.~D'Agostino {\it et al.},
  Nucl.\ Phys.\  A {\bf 749}, 55 (2005), [arXiv:nucl-ex/9906004].




\bibitem{SV1}
V.~V.~Skokov and D.~N.~Voskresensky, arXiv:0811.3868 [nucl-th],
JETP Letters {\bf 90}, 245 (2009).
\bibitem{SV2}
V.~V.~Skokov and D.~N.~Voskresensky,  Nucl. Phys. {\bf A828}, 401
(2009).

\bibitem{RandrupGQP} J.~Randrup, arXiv:0903.4736
[nucl-th], Phys. Rev. {\bf C79}, 054911 (2009).
\bibitem{SV3}V.~V.~Skokov and D.~N.~Voskresensky,  Nucl. Phys. {\bf A847},
253 (2010).
\bibitem{R10}
 J. Randrup,  Phys. Rev. {\bf C82},  034902 (2010).
\bibitem{PS}
A.~Z. Patashinsky and B.~I. Shumilo, JETP {\bf 50}, 712 (1979).



\end{thebibliography}
\end{document}